\documentclass[journal]{IEEEtran}
\usepackage{graphicx}
\usepackage{epsf}
\usepackage{mathtools}
\usepackage{amssymb}
\usepackage{amsmath, amsfonts}
\usepackage{latexsym}
\usepackage{multirow}
\usepackage{multicol}
\usepackage[table]{xcolor}
\usepackage{color}

\usepackage{tabularx}
\usepackage{lipsum}

\usepackage{mathrsfs}

\usepackage{fixltx2e}

\usepackage[mathscr]{euscript}

\usepackage{commath}
\usepackage{MnSymbol}

\usepackage{subcaption}

\usepackage{mathtools} 
\DeclarePairedDelimiterX\setc[2]{[}{]}{\,#1 \;\delimsize\vert\; #2\,}
\DeclarePairedDelimiterX\parth[2]{(}{)}{\,#1 \;\delimsize\vert\; #2\,}

\DeclareMathOperator*{\argmax}{argmax}

\PassOptionsToPackage{hyphens}{url}
\usepackage[unicode=true, bookmarks=true, bookmarksnumbered=true, bookmarksopen=true, bookmarksopenlevel=1, breaklinks=false, pdfborder={0 0 0}, pdfborderstyle={}, backref=false, colorlinks=false]{hyperref}

\usepackage[ruled,linesnumbered]{algorithm2e}

\usepackage{theorem}
{
\theorembodyfont{\rmfamily}

\newtheorem{remark}{Remark}
\newtheorem{lemma}{Lemma}

\newtheorem{definition}{Definition}

}

\usepackage{dsfont}

\usepackage{tikz}
\newcommand*\circled[1]{\tikz[baseline=(char.base)]{
\node[shape=circle,draw,inner sep=0.6pt] (char) {#1};}}

\definecolor{orange}{RGB}{255,127,0}
\definecolor{blue}{RGB}{0,0,255}
\definecolor{red}{RGB}{255,0,0}
\definecolor{green}{RGB}{50,160,50}
\definecolor{grey}{RGB}{125,120,125}

\begin{document}
{
\title{\fontsize{17}{21}\selectfont Byzantine-Fault-Tolerant Consensus via Reinforcement Learning for Permissioned Blockchain-Empowered V2X Network}

\author
{
Seungmo Kim, \textit{Member}, \textit{IEEE}, and Ahmed S. Ibrahim, \textit{Member}, \textit{IEEE}

\vspace{-0.3 in}

\thanks{S. Kim is with the Department of Electrical and Computer Engineering, Georgia Southern University in Statesboro, GA, USA. Ahmed S. Ibrahim is with the Department of Electrical and Computer Engineering, Florida International University in Miami, FL, USA.}
}

\maketitle
\begin{abstract}
Blockchain has been at the center of various trust-promoting applications for vehicle-to-everything (V2X) networks. Recently, permissioned blockchains gain practical popularity thanks to their improved scalability and diverse needs for different organizations. One representative example of permissioned blockchain is Hyperledger Fabric. Due to its unique execute-order procedure, there is a critical need for a client to select an optimal number of peers. The interesting problem that this paper targets to address is the tradeoff in the number of peers: a too large number will lead to a lower scalability and a too small number will leave a narrow margin in the number of peers sufficing the Byzantine fault tolerance (BFT). This channel selection issue gets especially challenging to deal with in V2X networks due to the mobility: a transaction must be executed and the associated block must be committed before the vehicle leaves a network. To this end, this paper proposes an optimal channel selection mechanism based on reinforcement learning (RL) to keep a Hyperledger Fabric-empowered V2X network impervious to dynamicity due to mobility. We model the RL as a contextual multi-armed bandit (MAB) problem. The results prove the outperformance of the proposed scheme.
\end{abstract}


\section{Introduction}\label{sec_intro}
\subsection{Background}
Vehicle-to-Everything (V2X) communications are acknowledged to have a massive potential to significantly decrease the number of vehicle crashes, thereby reducing the number of associated fatalities \cite{gdot_1}. The capability gave V2X communications the central role in constitution of intelligent transportation system (ITS) for connected and autonomous vehicles (CAVs).

Meanwhile, the blockchain technology has been gaining widespread interest based on its capability of providing secure, access-regulated interactions and transactions. However, to be applied in V2X networks, the key challenge lies in keeping the performance of a consensus due to the networks' dynamicity attributed to mobility \cite{lett19}\cite{access19}. In general, a consensus algorithm is defined as a process to achieve agreement on a single data value among distributed processes or systems. Consensus algorithms are designed to achieve a certain degree of reliability even in a network involving unreliable nodes.

Permissioned blockchains are getting popular as a means to address this issue. In many distributed blockchains, such as Ethereum and Bitcoin, which are not permissioned (also known as ``public''), any node can participate in the consensus process, wherein transactions are ordered and bundled into blocks. Because of this characteristic, these systems rely on \textit{probabilistic consensus} algorithms, which eventually guarantee ledger consistency to a high degree of probability, but which are still vulnerable to divergent ledgers (also known as a ledger ``fork''), where different participants in the network have a different view of the accepted order of transactions. Permissioned blockchains work differently. They aim at a \textit{deterministic consensus} among all the nodes participating in a validation process.

\subsection{Hyperledger Fabric}
The Hyperledger Fabric (``Fabric'' from now) has the widest popularity these days owing to its design as modular consensus protocols, which allows the system to be tailored to particular use cases and trust models. It features a node called an orderer (also known as an ``ordering node'') that does this transaction ordering, which, along with other orderer nodes, forms an ordering service. Because Fabric's design relies on algorithms, any block validated by the peer is guaranteed to be final and correct. Ledgers cannot fork the way they do in many other distributed and permissionless blockchain networks.

Also, the Fabric employs an \textit{execute-order} architecture, which requires all peers to execute every transaction and all transactions to be deterministic. Conversely, existing blockchain systems employ the opposite ``order-execute'' architecture: examples range from public blockchain such as Ethereum to permissioned ones that usually employ a consensus mechanism based on crash fault tolerant (CFT) or Byzantine fault tolerant (BFT) \cite{fabric_ibm18}. The limitation of the order-execute architecture is apparent: every peer executes every transaction and transactions must be deterministic.

In a Fabric network, the scalability is predominantly determined by the complexity of its endorsement policy \cite{throughput18} and an ordering service where a consensus has to be reached \cite{latency_ibm18}. Specifically, the validation of a transaction's endorsements requires evaluation of endorsement policy expression against the collected endorsements and checking for satisfiability \cite{throughput18_24}, which is usually achieved via a \textit{gossip protocol} in a BFT-based consensus mechanism. This is the key bottleneck in the scalability: a larger number of peers participating in validation usually causes a longer latency and hence a lower throughput.

Moreover, there is a pitfall in the Fabric's execute-order structure \cite{ibm_arxiv19}. Since an application is executed before validation of the associated transaction, the key drawback of this system occurs when the transaction turns out invalid at the end. It incurs a security problem and also waste of resources executing the application not complying the endorsement policy.

\subsection{Reinforcement Learning for Performance Optimization}
In this paper, we propose to apply reinforcement learning (RL) to optimize the selection of a channel in a Fabric network implemented in a V2X environment. However, there still remain challenges to address. Specifically, the learning is extremely complicated due to the dynamicity, which necessitates that the learning framework itself must be resilient and flexible according to the environment.

This paper proposes a learning mechanism formulated as a multi-armed bandit (MAB) problem, which enables a vehicle, without any assistance from an external infrastructure, to autonomously learn the environment and adapt its channel access behavior according to the outcome of the learning.

The MAB simplifies a RL by removing the learning dependency on \textit{state} and thus providing evaluative feedback that depends entirely on the \textit{actions}. The actions usually are decided upon in a greedy manner by updating the benefit estimates of performing each action independently from other actions. To consider the state in a bandit solution, \textit{contextual bandits} may be used \cite{RLbandit19_chu}. In many cases, a bandit solution may perform as well as a more complicated RL solution or even better. Many bandit algorithms feature stronger theoretical guarantees on their performance even under adversarial settings \cite{RLbandit19}.

\textit{Thompson sampling (TS)} (also known as \textit{posterior sampling}) provides a statistically efficient approach that tackles the exploration-exploitation dilemma by maintaining a posterior over models and choosing actions in proportion to the probability that they are optimal \cite{googlebrain18}. We will show in this paper that the endorsing peer selection problem can be solved via Thompson sampling.

\subsection{Contribution of This Paper}\label{sec_intro_contribution}
This paper proposes an endorser selection mechanism based on RL that is performed autonomously by a client. Specifically, this paper features the following contributions:
\begin{itemize}
\item It is the first work investigating the feasibility of the Hyperledger Fabric for V2X networking.
\item It (i) adopts RL for accomplishing the aforementioned novel consensus protocol and (ii) models the optimization as a \textit{contextual MAB} problem as means to achieve RL. The prior work such as \cite{icc19_tii19} provided only little technical detail on the RL scheme itself, while focusing on the BFT protocol design. This paper takes a more \textit{balanced} view on both of the BFT and RL.
\item It provides a \textit{spatiotemporal} analysis framework for evaluating the performance of a blockchain system applied to a V2X network. The framework has advantages on several fronts: (i) the dynamics of vehicles are modeled by using stochastic processes; (ii) the time effects on the blockchain performance are evaluated; and (iii) the performance of RL is evaluated as Bayesian statistics.
\end{itemize}

\section{System Model}\label{sec_system_model}
\subsection{Blockchain}
This paper assumes a V2X network on which a permissioned blockchain is formed based on the \textit{Hyperledger Fabric v2.0} \cite{fabricv20}. Specifically, the roadside units (RSUs) act as \textit{peers} that participate in endorsement and consensus (i.e., validation and commit) in the permissioned blockchain, while the onboard units (OBUs)---i.e., vehicles---are served as \textit{clients}. Applying the Fabric to this architecture, the RSUs have the authority to validate and order a block, which means that all the endorsing peers and orderers are selected from the RSUs. Meanwhile, OBUs are treated as clients of the execution and ordering services. By this architecture, we mean to make the blockchain system operate stably despite the vehicles' frequent entry into and departure from the blockchain network.

We notice that this architecture makes practical sense because a blockchain system will likely be managed by a certain party such as a state or federal organization or a private enterprise, through which vehicles pass and some of them may generate blocks that should be processed in the blockchain managed by such an organization.

Let us provide some technical backgrounds for the Fabric. The fourth box from the top named ``Fabric network'' in Fig. \ref{fig_structure} will aid understanding if referred to while reading this subsection.

A network is comprised primarily of a set of peer nodes (or, simply, peers). Peers are a fundamental element of a Fabric network because they host ledgers and smart contracts. Other major elements of a Fabric network include channels, orderers, and organizations. A \textit{channel} is defined as a mechanism via which peers interact with each other and with applications and transact privately. To understand the concept of a channel in relation to a peer, a channel is a logical structure that is formed by a collection of peers: peers provide the control point for access to, and management of, channels. An \textit{organization} is a party that has ownership and thus control of a Fabric network. That is, a Fabric network can be built up from the peers owned and contributed by the different organizations. For instance, a certain network established in a physical area can be governed by multiple parties: e.g., city, county, state, federal, and private enterprise. The Fabric allows a set of physical resources shared by multiple parties while each of the parties can maintain a private network built upon the resources, which is called an organization.

The Raft is a protocol that the Fabric adopts for consensus in an ordering service. Raft uses a ``leader and follower'' model where a leader is dynamically elected among the ordering nodes in a channel, and that leader replicates messages to the follower nodes. Raft is considered a step forward to BFT from Kafka, its predecessor that was based on CFT.

Also, as a significant remark, we remind that from v2.0, via an ordering service named ``Raft,'' the Fabric started to provide a \textit{BFT-based} consensus for validation and commit of a block. We emphasize that this also suits to address the dynamicity of a V2X network, which is highly dynamic in the network topology and the member composition at a certain time, which implies a far higher possibility of any malice or fault. As such, the employment of Fabric is justified in both aspects of efficiency and security.

\begin{figure*}
\centering
\begin{subfigure}{.495\textwidth}
\centering
\includegraphics[width=\linewidth]{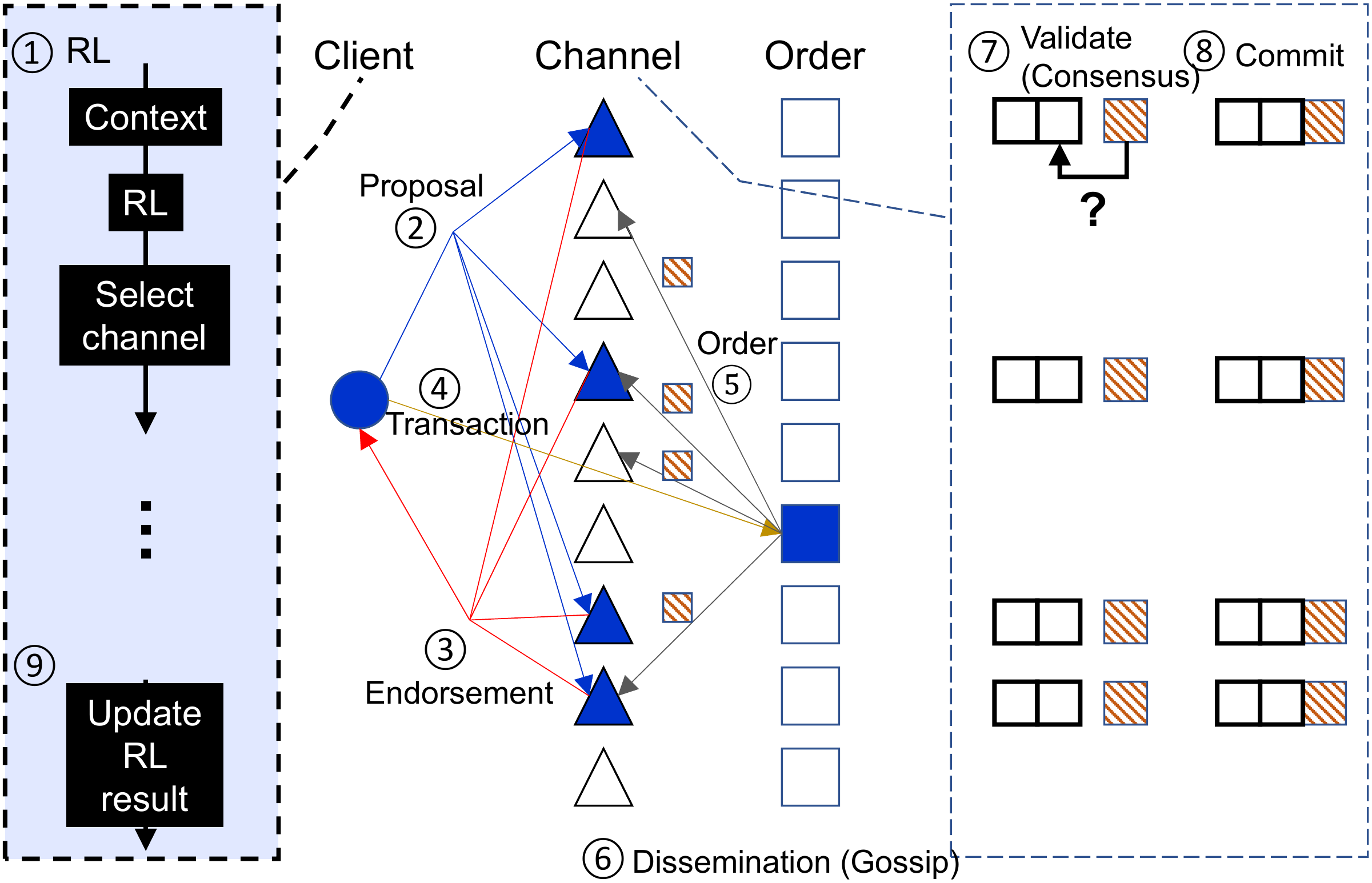}
\vspace{0.01 in}
\caption{Messaging process view}
\label{fig_architecture}
\end{subfigure}
\hfill
\begin{subfigure}{.495\textwidth}
\centering
\includegraphics[width=\linewidth]{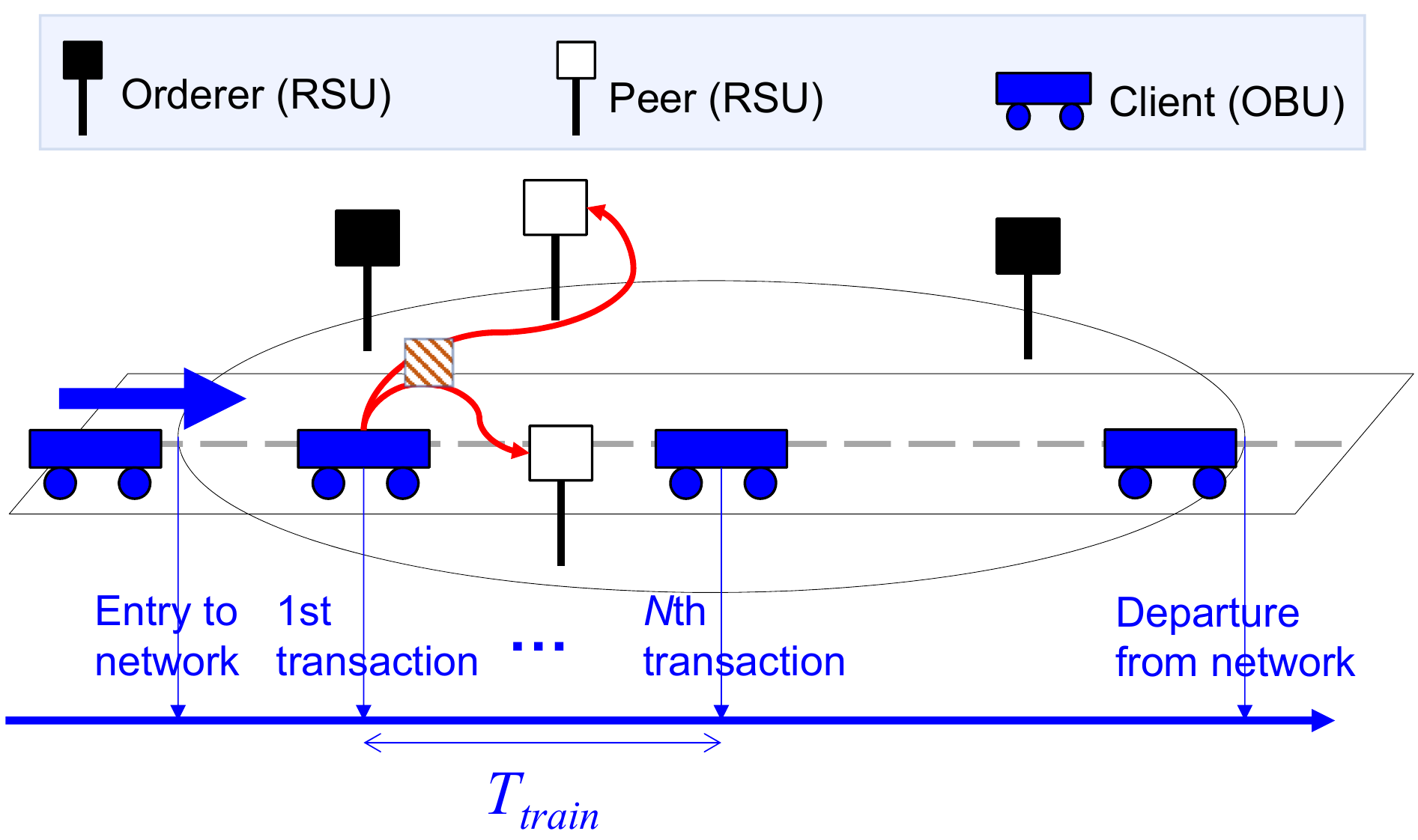}
\vspace{0.05 in}
\caption{Spatiotemporal view}
\label{fig_timing}
\end{subfigure}
\caption{Proposed endorser selection mechanism}
\label{fig_mechanism}
\end{figure*}

As shall be detailed in Section \ref{sec_proposed}, the key problem statement of this paper is based on the tradeoff of channel selection. By definition, \textit{channels} partition a Fabric network in such as way that only the stakeholders can view the transactions. In this way, organizations are able to utilize the same network while maintaining separation between multiple blockchains. The mechanism works by delegating transactions to different ledgers. Members of the particular channel can communicate and transact privately, while other members of the network cannot see the transactions on that channel. The Raft consensus service allows an orderer to \textit{select} a channel through which it will serve the ordering service. As such, this paper focuses on \textit{finding an optimal channel} that minimizes the latency and maximizes the throughput.

Next, we assume that not all RSUs are connected to each other. A RSU usually has no wired connection, which causes that it only has a finite coverage \cite{rsu}. Taking this practical aspect into consideration, we assume that only a certain number of RSUs falling into each other's communications range are connected. Interestingly, the Fabric does already consider this type of situation, which leads to employment of a \textit{Gossip protocol} in disseminating information to reach a consensus during a block validation procedure.

It is also noteworthy that we consider a \textit{discrete} time setting. Specifically, in each period $t = 1, \cdots , T$, where $T \in \mathbb{N}$ is a finite time horizon. It is a \textit{synchronous} network \cite{80211p}, wherein all the clients (i.e., vehicles) and peers (i.e., RSUs) refer to the same discrete time $t$. As such, in the evaluation of this network's performance, we measure at any arbitrary node (i.e., a vehicle or a RSU) the number of slots that are consumed to process a transaction to append a block to the chain. We also remind to assume the \textit{same length} of $t$ for all nodes.

\subsection{Geometry}
We reiterate that not all nodes are connected directly to each other; however, every node is equipped with communications functionality and hence is able to exchange a transaction or a block to each other.

This paper adopts the stochastic geometry for characterization of a V2X network on a space \cite{globecom18}-\cite{arxiv2005}. They commonly rely on the fact that uniform distributions of nodes on $X$ and $Y$ axes of a Cartesian-coordinate two-dimensional space yield a Poisson point process (PPP) on the number of nodes in the space. The distributions of RSUs and OBUs are modeled as an independent homogeneous PPP $\Phi_{r}$ and $\Phi_{o}$ with the vehicle density $\lambda_{r}$ and $\lambda_{o}$.

A two-dimensional space $\mathbb{R}^2$ is defined with the length and width of $l$ and $w$ meters (m), respectively. To capture a more dynamic and realistic movement of nodes in a vehicular network, this system model considers \textit{no separation of lanes}. Notice that such a generalized model enables the subsequent analyses more widely applicable \cite{access19}. Furthermore, to consider the most generic vehicle movement characteristic, this model assumes that every vehicle can move in any direction, which enables the system to capture every possible movement scenario including flight of unmanned aerial vehicles (UAVs), lane changing, intersection, and pedestrian walking.

\section{Proposed Mechanism}\label{sec_proposed}
As was introduced in Section \ref{sec_intro_contribution}, this paper proposes to enable a vehicle to (i) \textit{autonomously learn} about a channel that provides an optimal number of peers and (ii) hence minimize the latency and maximize the throughput.

\subsection{Improvement to the Hyperledger Fabric Architecture}
According to the Fabric's execute-order mechanism, a client works with only a certain subset of the peers (i.e., not all of the peers) depending on the endorsement policy in which it operates \cite{fabric_ibm18}. Unless certain peers are designated in the policy, the client \textit{randomly} selects the peers that will endorse its transaction. This is the part that this paper targets to improve: we propose a mechanism in which a client \textit{learns} to improve its selection of a channel.

Fig. \ref{fig_architecture} demonstrates the \textit{proposed RL-based execute-order mechanism} in a Fabric network. Specifics of the procedure is as follows: \circled{1} Each vehicle has a training done before the beginning of joining a network; \circled{2} When an application invokes, the vehicle sends a proposal to the selected number of endorsers; \circled{3} The endorsers send the result of execution back to the client; \circled{4} The client sends the endorsed transaction to an order; \circled{5} the orderer puts the transaction into a block along with other transactions and multicasts the block to a set of endorsers that are directly connected to it; \circled{6} The endorsers use a \textit{gossip protocol} to disseminate the block to all among themselves; \circled{7} The endorsers compare their ledgers if they are all final and validate the new block; \circled{8} Once the endorsers reach a \textit{consensus}, they append the block to the chain.

\subsection{RL for Channel Selection}
Now, let us take a closer look at the RL components in the proposed framework.

\subsubsection{Spatiotemporal View}
It is critical for a vehicle to collect the \textit{prior} distribution for each channel Fig. \ref{fig_timing} illustrates a spatiotemporal view on a vehicle from its first entry into a Hyperledger Fabric network to departure. Before entry, the vehicle sends a Join Request (REQ) to the closest RSU, from which it receives a Join Confirm (CFM) upon entry to the network. The Join CFM message contains the information that is necessary to train itself: i.e., the minimum required number of endorsers and the latest number of clients queued in each endorser.

An arbitrary vehicle $i$ is designed to spend a certain length of time, $T_{\text{train}}$, observe the context $\mathbf{c}_{i}$ and update the reward $\mathbf{r}_{i}$. After $T_{\text{train}}$ elapses, the vehicle exploits the learned rewards among the arms.

\subsubsection{Problem Formulation}\label{sec_proposed_problem}
We model the RL of the proposed framework as a MAB problem. Therein, in time slot $t$, vehicle $i$ (i.e., a client for a blockchain) becomes an agent that observes the context and chooses an action based on the reward achieved from the action. The MAB problem is formulated as follows.

Further, we characterize the proposed framework as a \textit{contextual MAB problem}. Since a primary RSU $i$ (the bandit) does not know the optimal action to take for a context a priori, the primary learns which action to select according to the context and hence becomes able to optimize the throughput. In order to learn the policy, the agent has to try out different arms (i.e., (i) the number and (ii) the IDs of voting peers) for on different contexts over time, which forms a contextual MAB problem.

By definition, a bandit problem is defined as a single state version of a Markov decision process (MDP). However, in our proposed system, the state of the agent (i.e., a vehicle) does not change after taking a certain action. For instance, suppose that a vehicle enters a network governed by the Fabric blockchain. Although we the mobility is a key factor distinguishing this work from other Hyperledger Fabric-based framework, as described in Fig. \ref{fig_timing}, the mobility can be translated down to a single variable, $T_{\text{dwell}}$. By this, the proposed framework can be modeled with the \textit{state} not changing for a vehicle in taking an action (i.e., arm). It means that the only factor affecting the agent's action is the context, $T_{\text{dwell}}$, since it represents other influencing factors such as the vehicle's position in relation to the network's coverage, the vehicle's speed, etc.

Therefore, we model our problem as a contextual MAB problem since, in this way, the vehicle does not simply learn which channel is the optimal on average, but instead it exploits additional information about other channels under a given traffic situation.

As such, in the proposed MAB problem, a newly entering vehicle (i.e., a client from the blockchain's point of views) is regarded a \textit{bandit}, and each channel is modeled as an \textit{arm} of the bandit.

Now, we formulate our proposed mechanism into an optimization problem. Let $x_{i,t} \in \mathbb{R}^1$ denote the context of vehicle $i$ at time $t$. Also, we denote by $\mathsf{Y}_{i,t} \in \mathbb{R}^2$ a vector of possible actions. Notice that $\tilde{y}_{i,t} \in \mathsf{Y}_{i,t}$ is an action selected by the policy $\pi\left(y_{i,t}|x_{i,t}\right)$. Now, the goal is to train $\pi$ in order to maximize the reward $r\left(y_{i,t} \hspace{0.03 in} | \hspace{0.03 in} x_{i,t}\right)$ over a training period $T$ via a finite-horizon decision problem, aiming to find the optimal $\pi^{*}$, which yields an optimal action $\left(y_{i,t}\right)^{*}$. We formulate this process of predicting the optimal $\pi^{*}$, which is formally written as
\begin{align}\label{eq_problem}
\left(y_{i,t}\right)^{*} = \pi^{*}\left(y_{i,t} \hspace{0.03 in} | \hspace{0.03 in} x_{i,t}\right) &= \argmax_{y_{i,t} \in \mathsf{Y}_{i,t}} \hspace{0.05 in} r\left( y_{i,t} \hspace{0.03 in} | \hspace{0.03 in} x_{i,t} \right)\\
&\text{subject to } c\left( y_{i,t} \hspace{0.03 in} | \hspace{0.03 in} x_{i,t} \right) \le C\nonumber
\end{align}
where $c(\cdot)$ denotes the cost and $C$ gives the maximum acceptable cost for operating action $y_{i,t}$ in context $x_{i,t}$. Notice that in this problem setting, we define the cost as the \textit{length of time taken for a consensus}, which is also called the \textit{latency} as shall be shown in Figure \ref{fig_latency}.

\begin{remark}\label{remark_KP}
(0-1 knapsack problem). \textit{We notice that the problem presented in (\ref{eq_problem}) is a \textit{0-1 knapsack problem (KP)} \cite{career_38}, which aims to maximize the reward while keeping the cost under a certain level. The key challenge is that a KP already is a non-deterministic polynomial-time (NP)-complete problem \cite{career_39}, which will make prediction of $\pi$ cumbersome even with input variable $X$ in a low dimension. As a means to deal with this challenge, we take a numerical approach to produce the results, which will be presented in Section \ref{sec_results}.}
\end{remark}

The key challenge in a \textit{MAB problem} lies in solving the \textit{exploration vs. exploitation dilemma}, since all actions should be explored sufficiently often to learn their rewards, but also those actions which have already yielded high rewards should be exploited \cite{tutorial_stanford}. Since we model our problem as a MAB, each vehicle needs to identify the best channel by carefully selecting one minimizing the latency and maximizing the throughput. The additional challenge in a \textit{contextual} MAB problem is how to exploit historical \textit{reward} observations under similar contexts. More technically, the problem of selecting an optimal channel comes from a tradeoff described in the following lemma:

\begin{lemma}\label{lemma_tradeoff}
(Tradeoff on the number of peers). \textit{Regarding the constraint in (\ref{eq_problem}), for a client, a tradeoff is formed in selecting a channel through which a transaction is executed and committed. In particular, the latency and throughput depend on ``the number of peers.'' If there are too many peers, a higher latency will be caused for endorsement and consensus; on the other hand, too few peers will more easily cause a consensus failure when Byzantine faults occur.}
\end{lemma}

\begin{algorithm}[hbtp]
\SetAlgoLined
Input: $T_{\text{train}}$\\
Initialize: $\mathbf{r}_{i}, \mathbf{y}_{i}$

\For{t = 1, $\cdots$, $T_{\text{dwell}}$}{
Send Join REQ and receive Join CFM;\\
\eIf{$t \le T_{\text{train}}$}{
\%--- \textit{Training} ---\%\\
$y_{i} \longleftarrow y_{i,t}$;\\
$\mathbf{r_{i}} \longleftarrow r_{i,t,k}$\\
\hspace{0.1 in} where $r_{i,k}\hspace{-0.02 in}\sim\hspace{-0.02 in}\text{Beta}\left(\alpha_{k},\beta_{k}\right) \longleftarrow \left(\alpha_{k},\beta_{k}\right)_{t}$\\\hspace{1.5 in} $\forall k \in \{1, \cdots, \mathsf{N}_{\text{arm}}\}$;\\
}{
\%--- \textit{Step \circled{1}: Channel selection} ---\%\\
\eIf{$\epsilon$-greedy}{
\eIf{rand $\le \epsilon$}{
\% \textit{Explore}\\
$\hat{k}_{i,t} = \mathcal{U}(\mathsf{N}_{\text{peer, min}}, \mathsf{N}_{\text{peer, max}}$);\\
}{
\% \textit{Exploit}\\
$\hat{k}_{i,t} = \text{argmax}_{k} \hspace{0.02 in} r_{i,k} \big|_{1, 2, \cdots, t-1}$;\\
}
}{
\%--- \textit{Thompson sampling} ---\%\\
$\hat{\theta}_{i,t} \sim \text{beta}\left(\alpha_{k},\beta_{k}\right)$ for $k = 1, \cdots, \mathsf{N}_{\text{arm}}$;\\
$\hat{k}_{i,t} \longleftarrow \max_{k} \mathbf{\hat{\theta}}_{i,t}$;
}
\%--- \textit{Steps \circled{2} and \circled{3}} ---\%\\
Send a transaction to the peers in channel $\hat{k}_{i,t}$;\\
Receive endorsement result from the peers in channel $\hat{k}_{i,t}$;\\
\eIf{Endorsement successful}{
\% Step \circled{4}\\
Request order;\\
(Steps \circled{5}-\circled{8} by orderers and endorsers)\\
\eIf{Validation successful}{
$r^{\text{ec}}_{i,t} \longleftarrow 1$;
}{
$r^{\text{ec}}_{i,t} \longleftarrow 0$;
}
}{
$r^{\text{ec}}_{i,t} \longleftarrow 0$;
}
\%--- \textit{Latency examination} ---\%\\
\eIf{Latency $\le T_{\text{dwell}}$}{
$r^{\text{ld}}_{i,t} \longleftarrow 1$;
}{
$r^{\text{ld}}_{i,t} \longleftarrow 0$;
}
\% Reward\\
$r_{i,t} \longleftarrow r^{\text{ec}}_{i,t} \cap r^{\text{ld}}_{i,t}$;\\
\% Step \circled{9}\\
$\left(\alpha_{k},\beta_{k}\right) \longleftarrow \left(\alpha_{k},\beta_{k}\right)_{t} + r_{i,t}$;
}
}
\caption{Proposed RL-based execution-order algorithm for a vehicle sending a block in Fabric-empowered V2X network}
\label{algorithm_rl}
\end{algorithm}

\subsubsection{Context}
Minimizing the need for modification to the current version of Fabric, we propose to design the \textit{context} as those can be defined within an endorsement policy.

\vspace{-0.1 in}

\begin{definition}\label{definition_contexts}
(Context: Client's dwelling time in a Fabric network). \textit{A client (i.e., a bandit in the MAB) makes an action based on its dwelling time in the Fabric blockchain network, which is denoted by $T_{\text{dwell}}$. The geographic information (e.g., the estimated radius of the network's boundary) is provided from the network via an endorsement policy in Join CFM upon joining of the network. Based on this information, each vehicle estimates its $T_{\text{dwell}}$ and uses as the context to make the selection on a channel. It is formally written as $T_{\text{dwell}} = r / v$ where $r$ gives the radius of a Fabric network and $v$ denotes the speed of the tagged vehicle.}
\end{definition}

\subsubsection{Reward}
We characterize this MAB as a ``Beta-Bernoulli bandit'' where the reward measured by vehicle $i$ in time $t$, $r_{i,t}$, is modeled to be either 1 (i.e., a success) or 0 (i.e., a failure).

\vspace{-0.1 in}

\begin{definition}\label{definition_reward}
(Reward: Beta-Bernoulli bandit). \textit{The reward for an action by client $i$ in time $t$ is defined as}
\begin{align}\label{eq_r}
r_{i,t} &= r^{\text{ec}}_{i,t} \cap r^{\text{ld}}_{i,t}
\end{align}
\textit{where $r^{\text{ec}}_{i,t} = \mathds{1}_{\text{ec}}$ and $r^{\text{ld}}_{i,t} = \mathds{1}_{\text{ld}}$. Notice that these indicator functions are associated with the following sets: $\mathcal{S}_{\text{ec}}$ contains transactions making through both execution and commit; and $\mathcal{S}_{\text{ld}}$ indicates transactions with a latency shorter than the client's dwelling time within the Fabric network.}
\end{definition}

Also, it is a \textit{Bayesian bandit} problem. It is required that some information on the \textit{prior distribution} is known to each bandit for most of the learning strategies such as $\epsilon$-greedy and TS \cite{ms_arxiv19}. However, a newly entering vehicle has \textit{no prior information} about the channels available in the Fabric network. This emphasizes the significance of a training period for the vehicle in order to make a decision that is close to an optimal.

The \textit{regret} of learning is defined as the difference between the reward achieved by vehicle $i$ in time slot $t$ and the optimal, which is formulated as
\begin{align}\label{eq_regret}
\rho_{i,t} = \left| r_{i,t} - r^{\ast}_{i,t} \right|
\end{align}
where $r^{\ast}_{i,t}$ denotes the reward that can be achieved by an optimal channel selection.

\subsubsection{Algorithm}
Now, we propose an \textit{online learning algorithm} implementing the proposed contextual MAB problem. Notice that the algorithm is meant to run at each vehicle \textit{on the fly} as the vehicle passes through an area operating a Fabric-based blockchain network. As described in Algorithm \ref{algorithm_rl}, the proposed framework features a RL mechanism to decide a channel to which it sends a transaction proposal.

As shown in Line 4, a vehicle is recognized by a Fabric network upon its entrance to the network, which is certified by receiving a Join CFM message from the network admin server.

As described in Line 6, vehicle $i$ should start a training period for $T_{\text{train}}$ slots upon new entry to a network. It observes context $c_{i,t}$ As a consequence, the vehicle collects the history of reward $r_{i}$ to update the \textit{prior} distribution for the reward from channel $k$. Specifically, after $T_{\text{train}}$ elapses, for each arm $k$, the vehicle piles \textit{successes} and \textit{failures} to the prior, which is characterized as $r_{i,k}\hspace{-0.02 in}\sim\hspace{-0.02 in}\text{Beta}\left(\alpha_{k}, \beta_{k}\right)$.

From Line 11, now the vehicle starts to utilize the learned prior distribution to select a channel when it needs to execute a transaction. As shown in Lines 13-23, a vehicle is able to choose between two representative strategies. In $\epsilon$-greedy, the vehicle still explores at the rate of $\epsilon$ and select channel $\hat{k}_{i,t}$ at random. At the other rate of $1 - \epsilon$, it selects the $k$ having the greatest mean reward so far. TS, on the other hand, performs a sampling for each of the $\mathsf{N}_{\text{arm}}$ arms and selects channel $\hat{k}_{i,t}$ as $k$ showing the largest sample. We compare the prediction performance between the two schemes Section \ref{sec_results}.

Line 27 implements that once the vehicle has selected a channel, it can send a proposal to peers belonging to the channel, $k_{i,t}$, whenever a transaction is generated by an application. Based on the Fabric, the peers execute the application and simulates the transaction if it is valid as per the endorsement policy. If valid, each peer sends the vehicle an endorsement as shown in Line 28.

\begin{table}[t]
\small
\caption{Parameters used for simulations}
\centering
\begin{tabular}{|p{0.5\linewidth}|p{0.4\linewidth}|}
\hline 
\textbf{Parameter}{\cellcolor{gray!20}} & \textbf{Setting} {\cellcolor{gray!20}}\\ \hline
\hline
V2X networking & DSRC\\
Blockchain system & Hyperledger Fabric v2.0\\
Consensus mechanism & BFT\\
Block dissemination for consensus & Gossip protocol\\
\# channels & 10\\
\# peers per channel & Random $\sim$ Uniform(5,10)\\
RL scheme & \{$\epsilon$-greedy, TS\}\\
\hline
\end{tabular}
\label{table_parameters}
\end{table}

Upon collection of a sufficient number of endorsements, the client now requests \textit{validation} of the transaction to the orderer, which Line 31 describes. The next step is to examine whether the execution and commit have been successful, i.e., whether $r^{\text{ec}}_{i,t} = 1$ or $0$, which is executed as in Lines 33 through 39. (We remind that this algorithm is designed to run on a vehicle; the tasks for orderers and endorsers---i.e., Tasks \circled{5}-\circled{8}---are written in parentheses in Line 32.)

Now, as Lines 41 through 46 show, the vehicle examines if it has been able to receive a reply from the order confirming \textit{commit} of the transaction while it still dwells in the network, the vehicle sets $r^{\text{ld}}_{i,t} = 1$, or $0$ otherwise.

Finally, the reward $r_{i,t}$ is computed and updated to the prior for each channel $k$, which are performed as Lines 47 through 50 in Algorithm \ref{algorithm_rl}.

\begin{figure}[t]
\centering
\includegraphics[width = \linewidth]{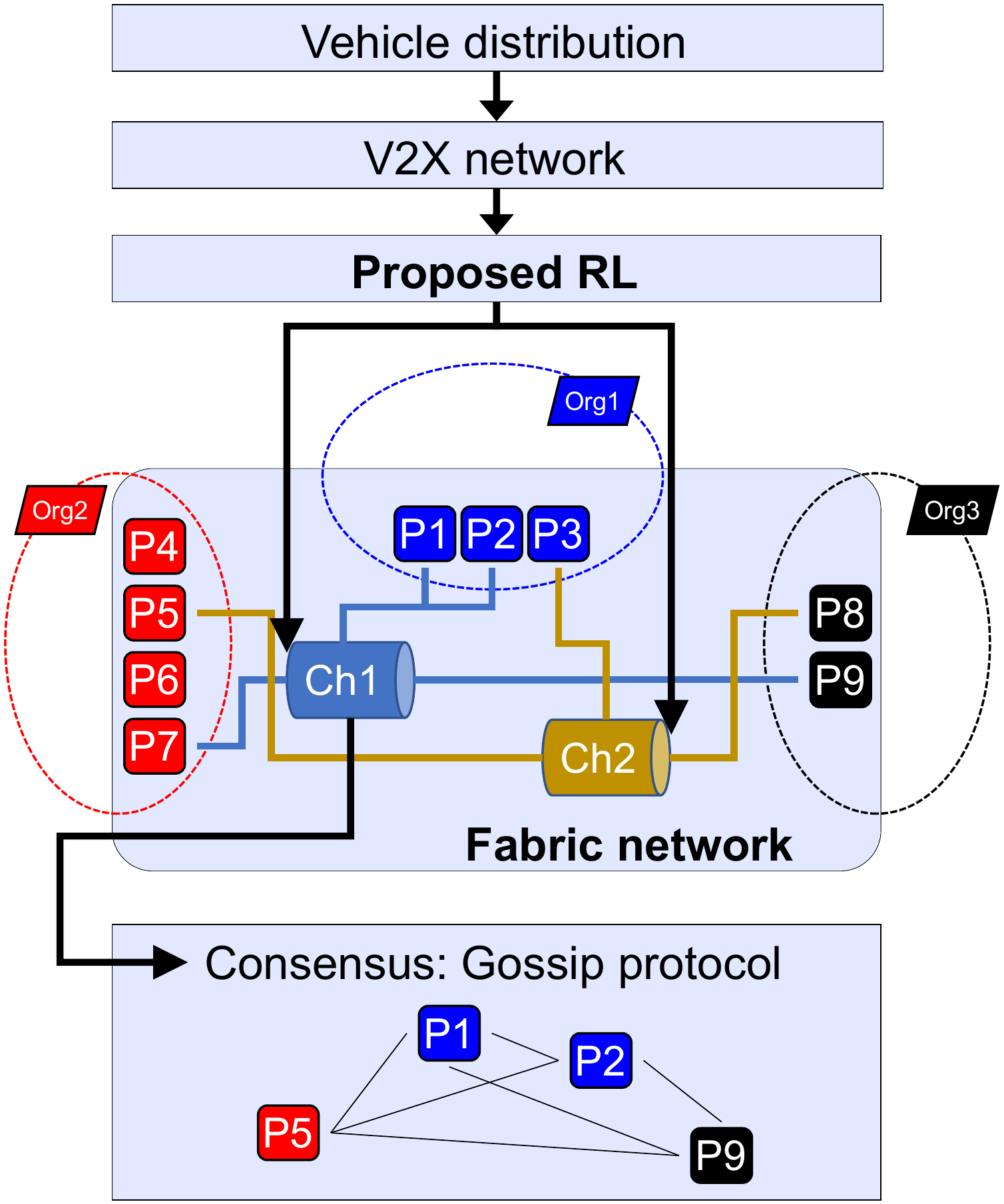}
\caption{The simulation software structure}
\label{fig_structure}
\end{figure}

\section{Performance Evaluation}\label{sec_results}
This section presents a detailed numerical evaluation of the proposed framework in terms of its performance in terms of convergence, regret, and scalability. For evaluating the performance, we constructed simulations for a Fabric-based V2X network on MATLAB.

\subsection{Setting}
Our test Fabric network consists of three organizations, each of which is with 5-10 endorsing peers for a total of 100 peer nodes. There are \{10, 20, 30\} channels established on different subsets of peers. There is one orderer node operating in Raft \cite{raft}. Table \ref{table_parameters} summarizes the parameters and their settings.

\subsection{Simulation Software Structure and Methodology}
Fig. \ref{fig_structure} illustrates key components consisting of the simulation software and their structure.

We found that simulation would serve the best efficiency as the main method to evaluate the performance of the proposed mechanism, based on several advantages \cite{fire08}.

First, as shall be presented through Figs. \ref{fig_heatmap} through \ref{fig_scalability}, the parameters' scope defining the network that operates the proposed mechanism is quite large, which makes it challenging to explore the parameters' dynamic orchestration in concert. Running simulations provides a relatively easier control over such a large space composed of various parameters with wide ranges of values. It gives an obvious advantage over mathematical derivations and testbed implementations. An example is the consensus process adopted by the Fabric. We assume a Gossip protocol as a means to achieve BFT consensus, which is usually complex to understand which aspect of the protocol dominates the overall performance. As an effort to circumvent such complexity, the proposed simulation methodology abstracts the Gossip protocol, which is illustrated in Fig. \ref{fig_structure}.

Second, simulations enable computations without being caught up with restrictions or errors caused by computing environmental factors including hardware, compiler, language, etc. Taking into account all the available options for all of those factors will complex the performance evaluation to a too high degree, which, as such, will make it hard to precisely identify the factors determining the key performance. In fact, existing literature has mentioned possible inaccuracy that could be caused by selection of a certain hardware \cite{intel19}.

\begin{figure}
\centering
\includegraphics[width = \linewidth]{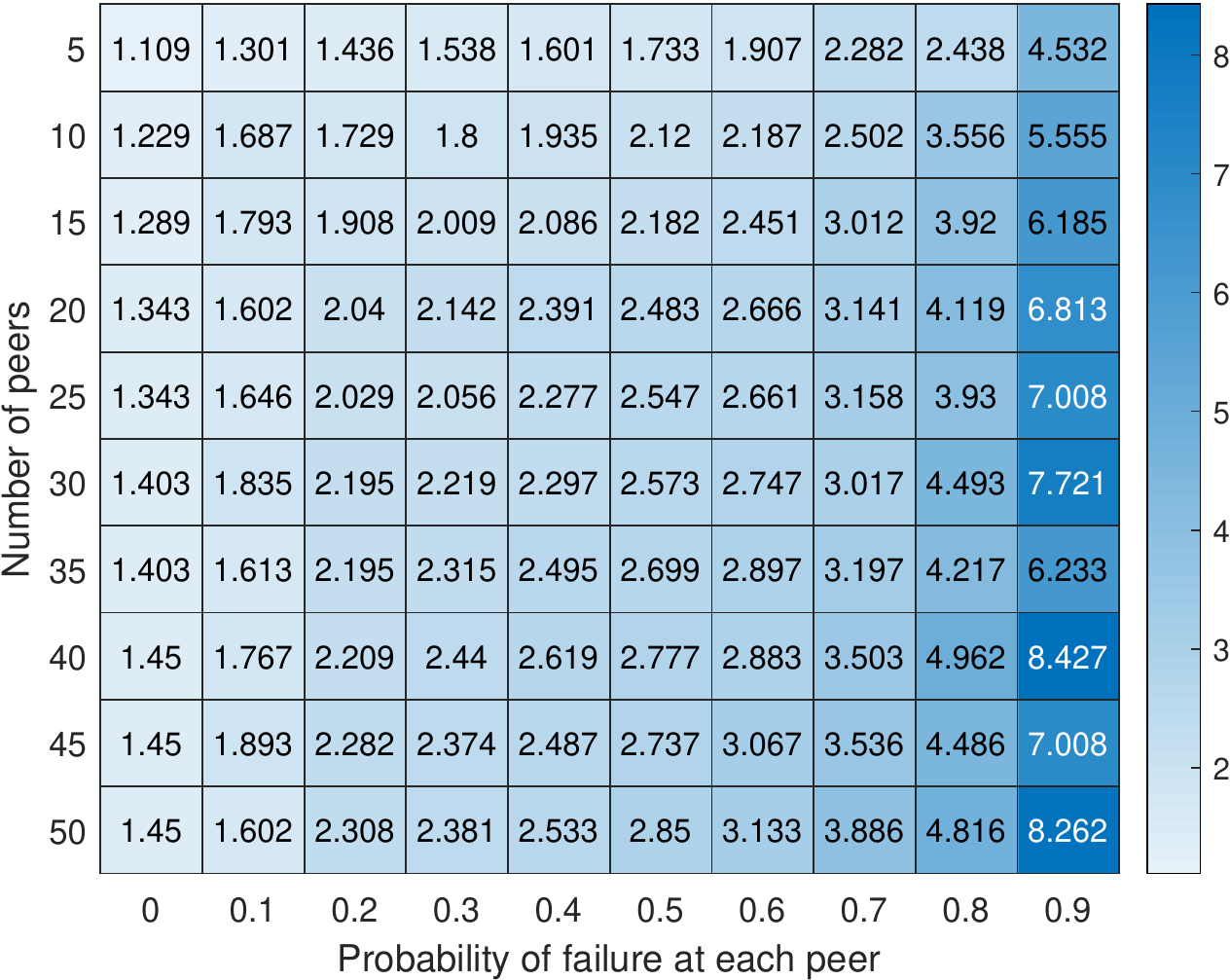}
\caption{Average latency (in seconds) versus \{number of peers, probability of failure\}}
\label{fig_heatmap}
\end{figure}

\subsection{Scenario}
Fig. \ref{fig_heatmap} shows the average latency versus (i) the number of peers in a channel and (ii) the probability of failure at each peer in a consensus procedure. It is obvious from the figure that a higher probability of failure at each peer incurs a higher latency. However, the pattern is less clear versus the number of peers. The reason is the tradeoff that was described in Lemma \ref{lemma_tradeoff}.

As such, we evaluate the performance in the following manner. A vehicle passes through the network for $T_{\text{dwell}}$ seconds. There are $\mathsf{N}_{\text{ch}} = 10$ channels in the Fabric network, each of which is assumed to have (i) a number of peers and (ii) a $p_{f}$ among those presented in Fig. \ref{fig_heatmap}. For instance, Channel 1 is expected to incur 2.308 seconds of latency with having 50 peers with $p_{f} = 0.2$. Channel 2, despite higher $p_{f} = 0.5$, it will give only 1.733 seconds of latency because of having 5 peers.

In the following results via Figs. \ref{fig_convergence} through \ref{fig_TPS}, the proposed mechanism will be shown to find the channel giving the minimum latency (and thus, the maximum throughput).

Exploiting the fact that we model the MAB problem as a Bernoulli-bandit, we evaluated two representative algorithms finding an optimal arm in a MAB problem. Fig. \ref{fig_convergence} shows the convergence performance of the two techniques. While $\epsilon$-greedy can focus on a proved arm at the rate of 90\% (since $\epsilon = 0.1$), it showed inefficiency by wasting time by still selecting irrelevant arms. On the other hand, TS is shown to better focus on the three successful arms as the learning progresses. In fact, TS has been evidenced to outperform other alternatives such as $\epsilon$-greedy and upper confidence bound (UCB) \cite{ms11}.

\begin{figure}
\centering
\includegraphics[width = \linewidth]{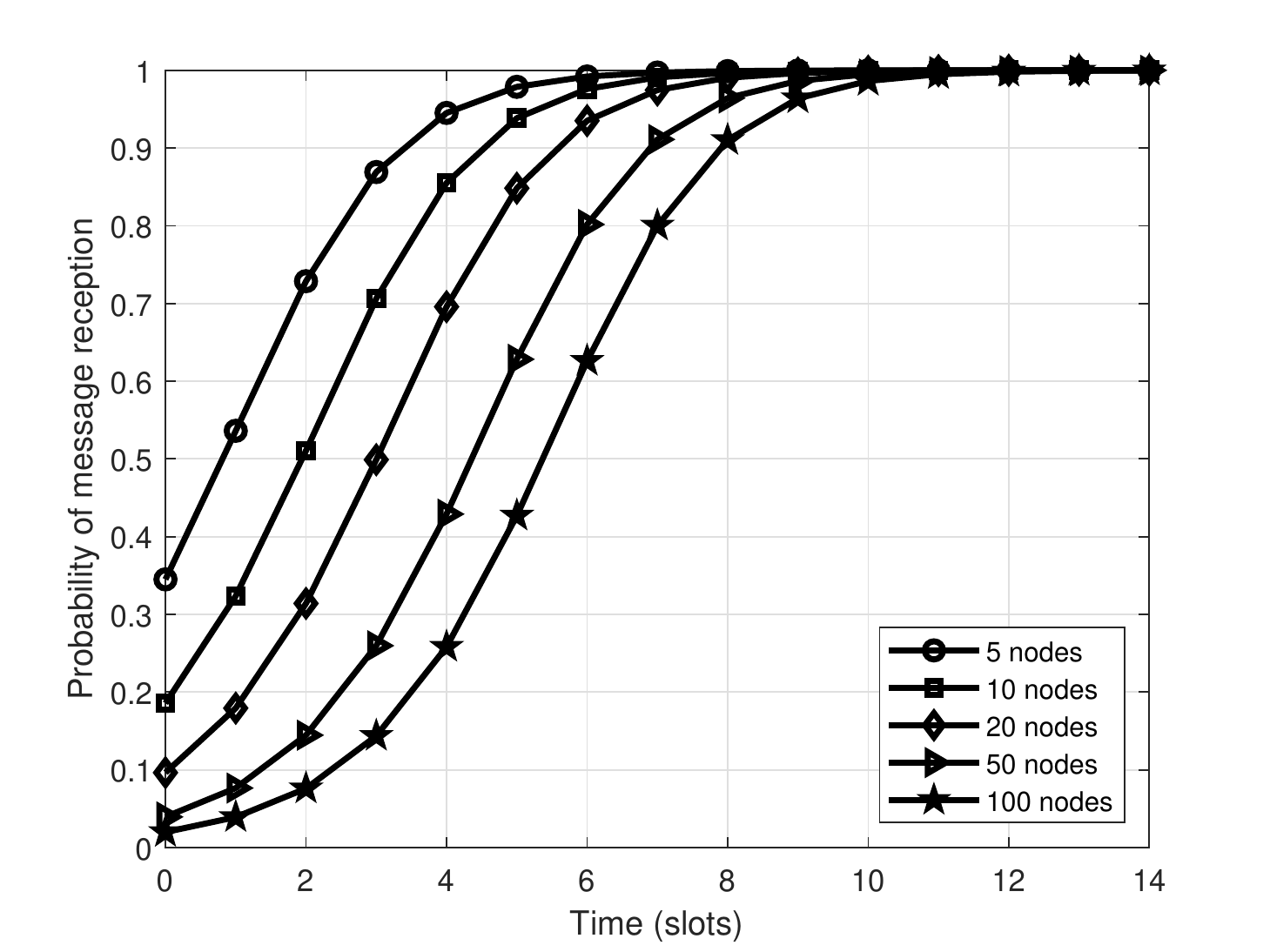}
\caption{Rate of dissemination of a message among a group of peers}
\label{fig_gossip}
\end{figure}

\subsection{Results}
Via Figs. \ref{fig_gossip} through \ref{fig_regret}, we demonstrate the performance of each component in the proposed framework.

\subsubsection{Consensus}
Figs. \ref{fig_gossip} shows the rate of dissemination of a block within a group of peers. Each curve indicates the length of time taken for propagation of a block from the master peer to all the other nodes. It is critical to recall that a consensus consumes the largest part in the latency for a block from being generated to being finally added to the blockchain. The dissemination rate ranges from 6 to 12 slots depending on the given numbers of peers--i.e., from 5 to 100.

To understand the result precisely, we remind that the Fabric adopts the \textit{Gossip} protocol for a consensus among the peers. We formulate a Gossip protocol as follows. Let $n$ denote the total number of nodes. Also, by $r_{t}$, we denote the proportion of nodes that have received the block sent from the source node after the execution of $t$ rounds. Meanwhile, $x_{t}$ gives the proportion of nodes that have not received the block yet: i.e., $x_{t} = 1 - r_{t}$. We assume that in the initial state, i.e., $t_{0}$, the source node has a block to commence a dissemination, which is given by $x_{0} = 1/n$ and $r_{0} = 1 - 1/n$. Now, for the propagation, we formulate the expectation of $x_{t}+1$ as a function of $x_{t}$ as follows. Assuming uniform random selection of a node to receive the propagation in the next time slot $t$, the expected rate of reception of the block by the randomly selected node can be written as \cite{gossip11}
\begin{align}\label{eq_gossip}
\mathbb{E}\left[r_{t+1}\right] = x_{t} \left( 1 - \frac{1}{n} \right)^{n\left(1 - x_{t}\right)}.
\end{align}
We remind that each curve in Fig. \ref{fig_gossip} describes $\mathbb{E}\left[r_{t+1}\right]$ versus $t$ for a value of $n$.

\begin{figure}
\centering
\begin{subfigure}{.495\textwidth}
\centering
\includegraphics[width=\linewidth]{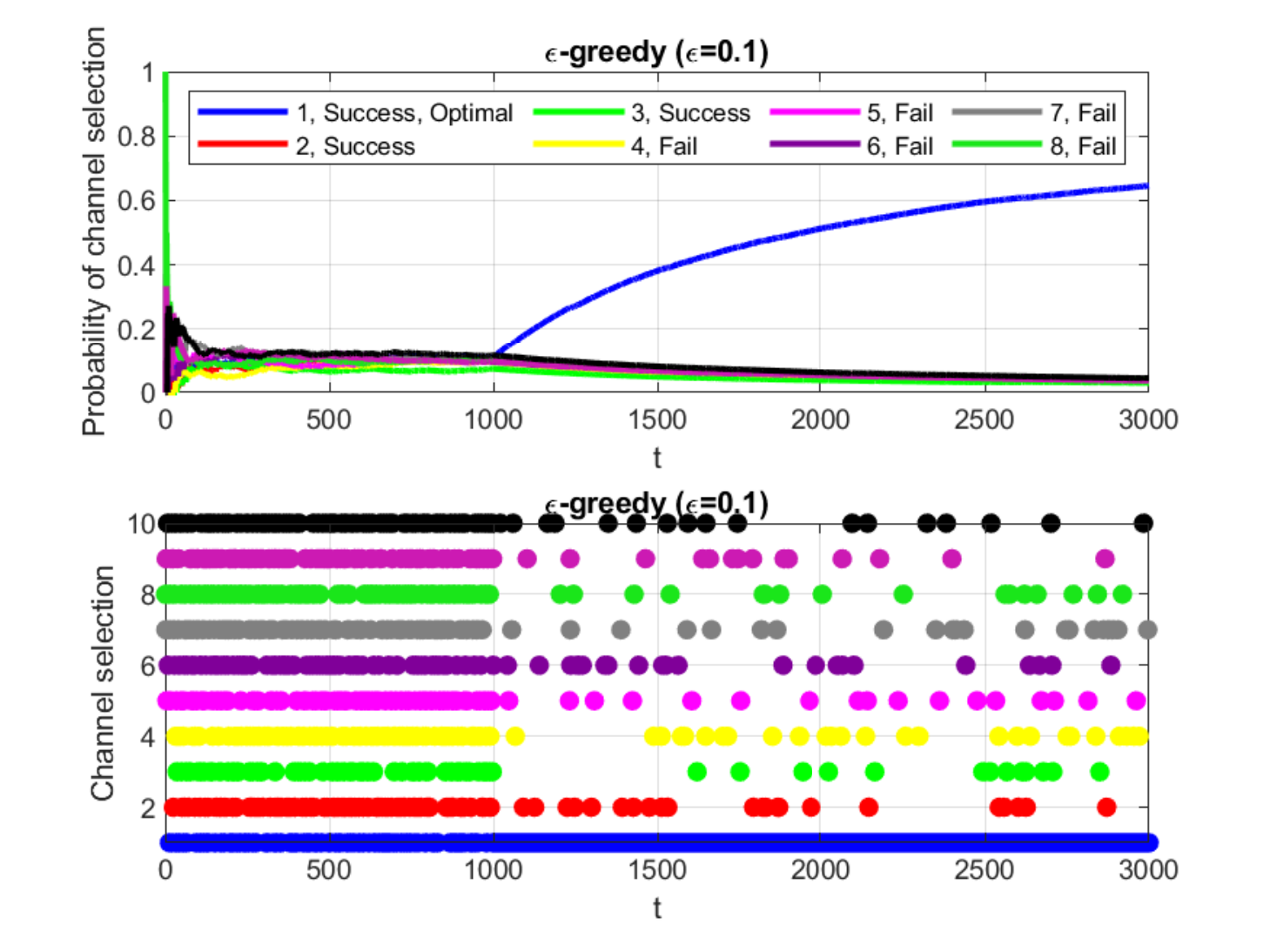}
\caption{$\epsilon$-greedy (With $\epsilon = 0 .1$)}
\label{fig_convergence_greedy}
\end{subfigure}
\hfill
\begin{subfigure}{.495\textwidth}
\centering
\includegraphics[width=\linewidth]{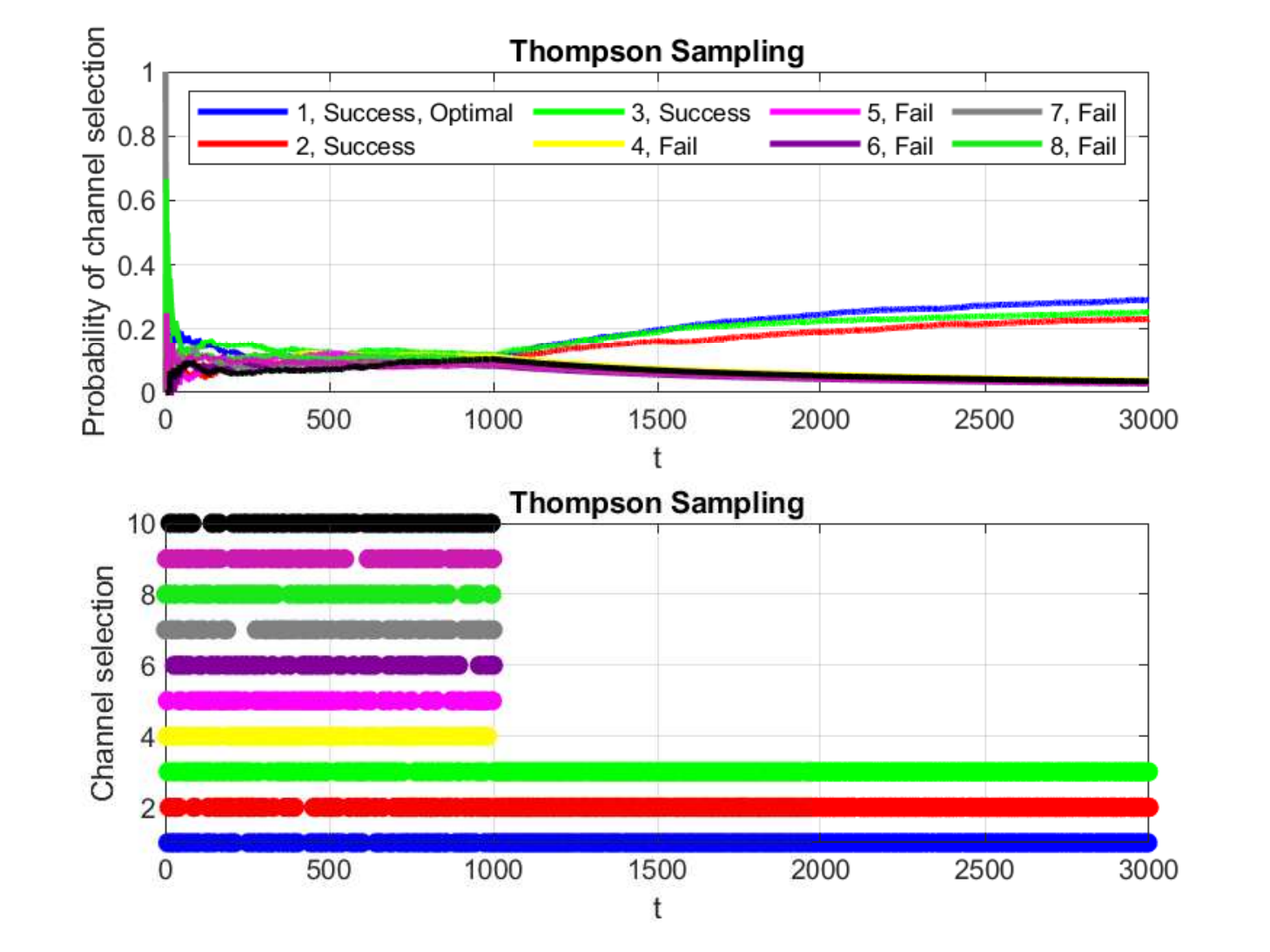}
\caption{TS}
\label{fig_convergence_TS}
\end{subfigure}
\caption{Convergence of the proposed RL algorithm (With 10 channels; For each subfigure: Upper: Selected channel at each $t$, Lower: Probability of each channel selection over $t$)}
\label{fig_convergence}
\end{figure}

\subsubsection{Time Convergence}
To evaluate the time complexity, we computed the average regret versus various lengths of $T_{\text{train}}$. Each of TS and $\epsilon$-greedy were run for $10^5$ iterations to demonstrate an average convergence performance. The results reveal the following four observations about the time complexity. First, TS shows a better concentration on the eligible channels, while $\epsilon$-greedy converges to a suboptimal, which is, however, still a success. Second, within each of TS and $\epsilon$-greedy, a larger number of $\mathsf{N}_{\text{arm}}$ was found to increase the regret, due to a larger load of searching, which can be translated as a higher time complexity. Third, commonly in the two techniques, a longer $T_{\text{train}}$ yielded a lower regret. Lastly, within $\epsilon$-greedy, a higher value for $\epsilon$ yields a lower regret since it has given a chance to learn from more explorations.

\subsubsection{Scalability}
Figs. \ref{fig_latency} and \ref{fig_TPS} show the scalability via the latency and throughput versus the number of clients, as a result of the proposed RL mechanism applied in the proposed Fabric system framework for V2X. Notice the definitions: \textit{throughput} is the rate at which transactions are committed to ledger, and \textit{latency} is the time taken from application sending the transaction proposal to the transaction commit.

The key observation is that the proposed mechanism (dotted lines) achieves a performance that is far closer to the optimal than the current Fabric's channel selection mechanism. The rationale is the proposed RL scheme enables a vehicle to select a channel that provides a close-to-optimal number of peers, addressing the tradeoff that was described in Section \ref{sec_proposed_problem}.

\begin{figure}
\centering
\begin{subfigure}{.495\textwidth}
\centering
\includegraphics[width = \linewidth]{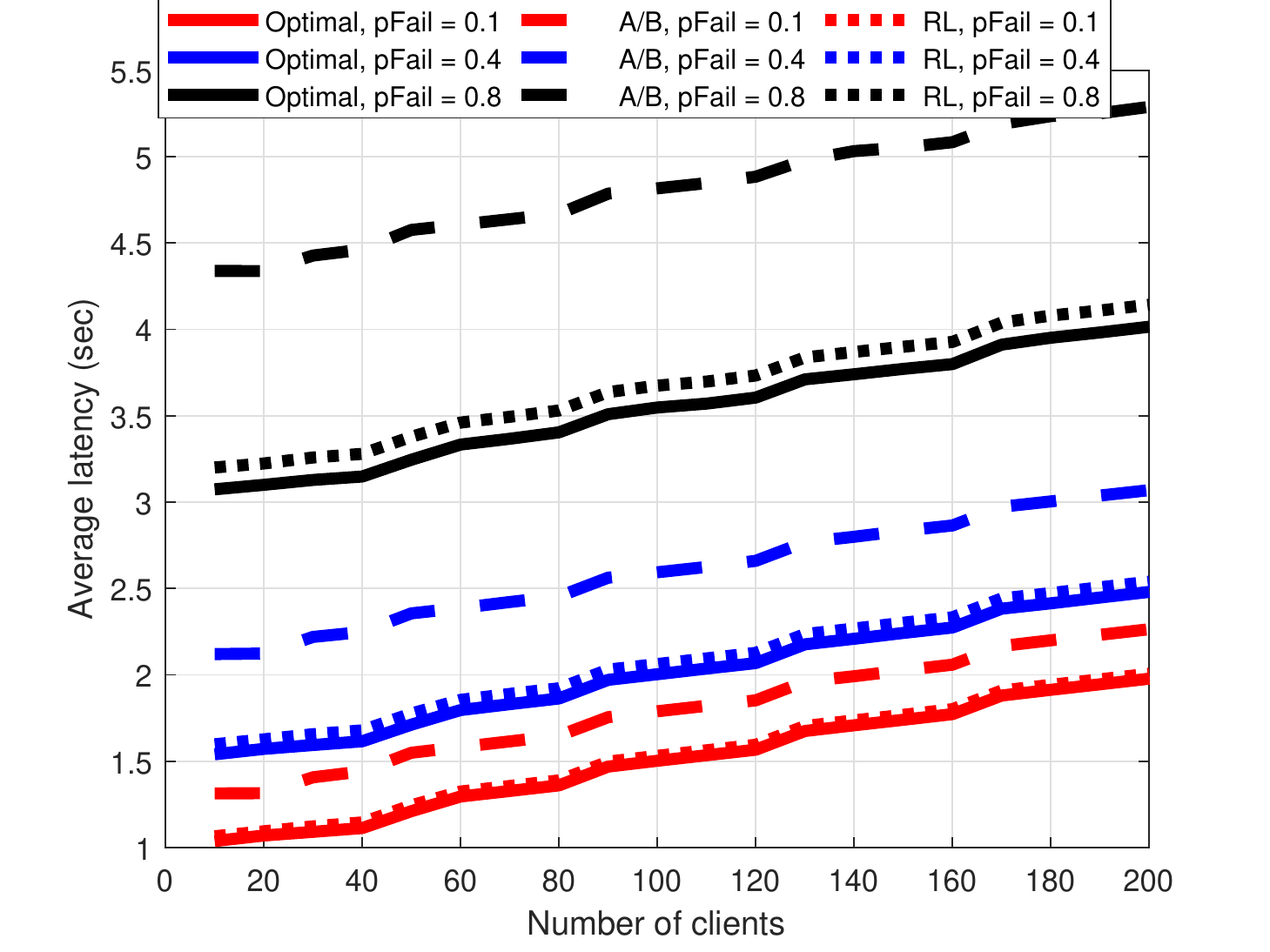}
\caption{Average latency vs. number of clients}
\label{fig_latency}
\end{subfigure}
\hfill
\begin{subfigure}{.495\textwidth}
\centering
\includegraphics[width = \linewidth]{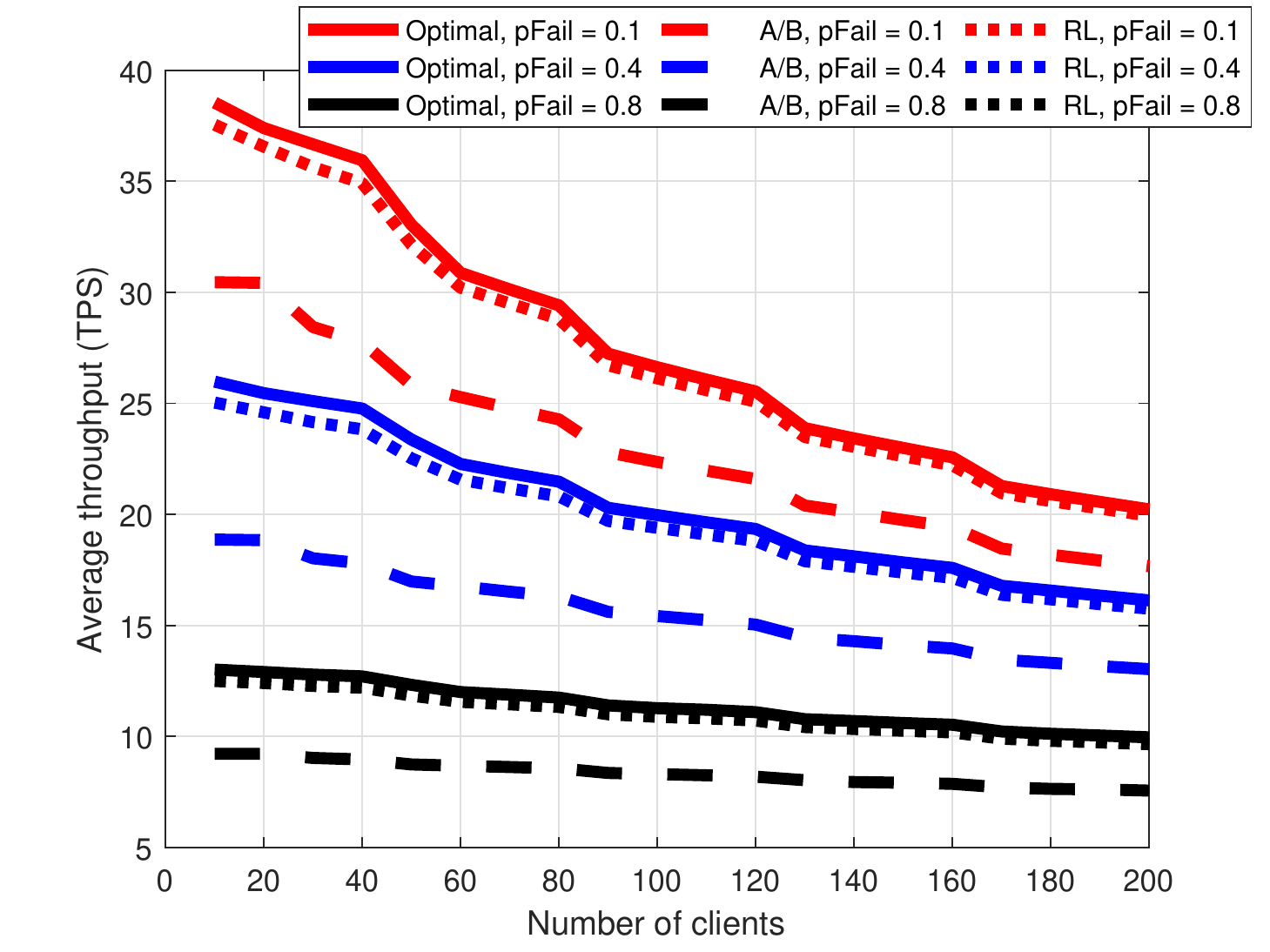}
\caption{Average throughput vs. number of clients}
\label{fig_TPS}
\end{subfigure}
\caption{Scalability}
\label{fig_scalability}
\end{figure}

\begin{figure}
\centering
\begin{subfigure}{.495\textwidth}
\centering
\includegraphics[width=\linewidth]{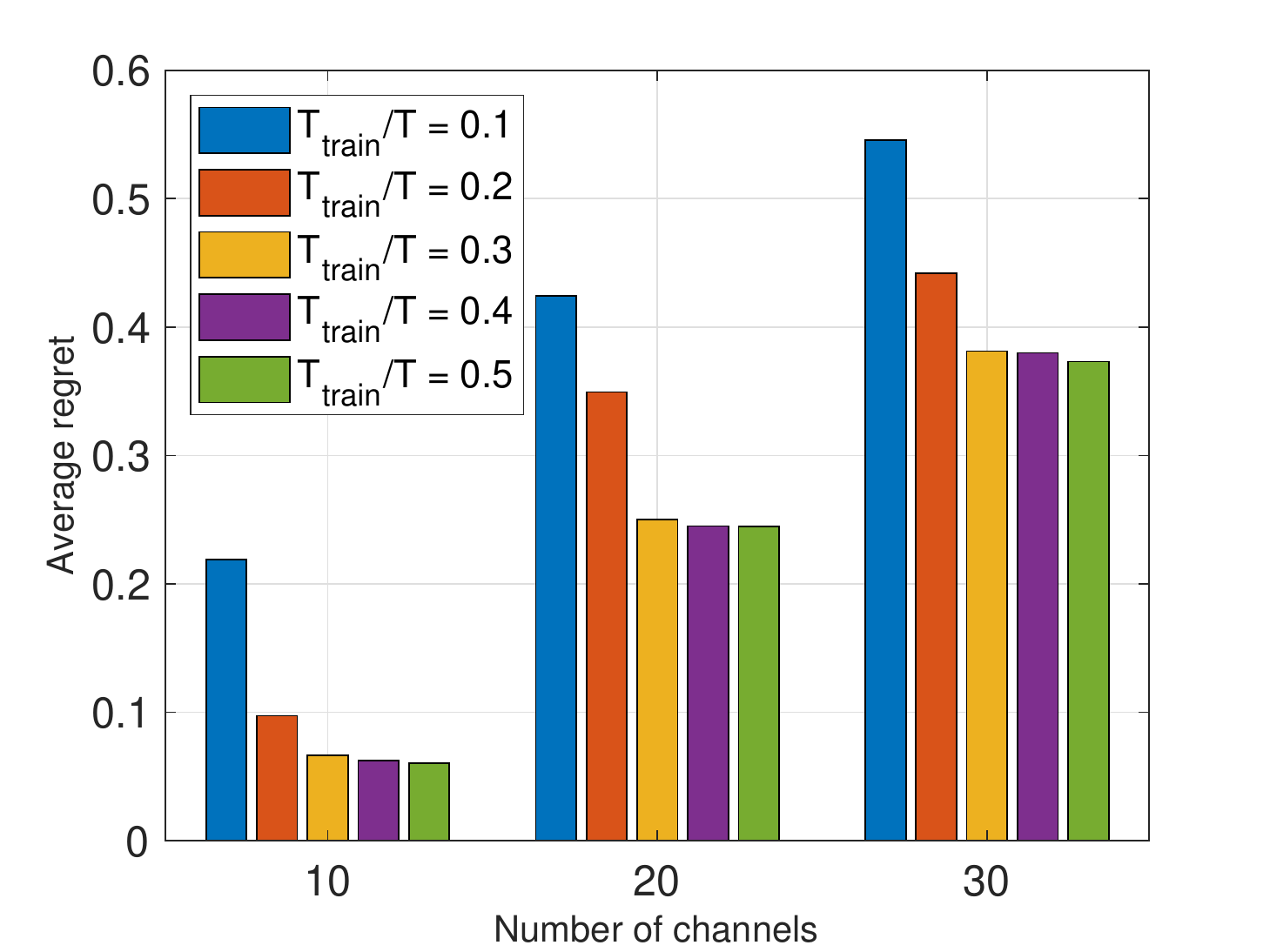}
\caption{$\epsilon$-greedy (With $\epsilon = 0 .1$)}
\label{fig_regret_greedy}
\end{subfigure}
\hfill
\begin{subfigure}{.495\textwidth}
\centering
\includegraphics[width=\linewidth]{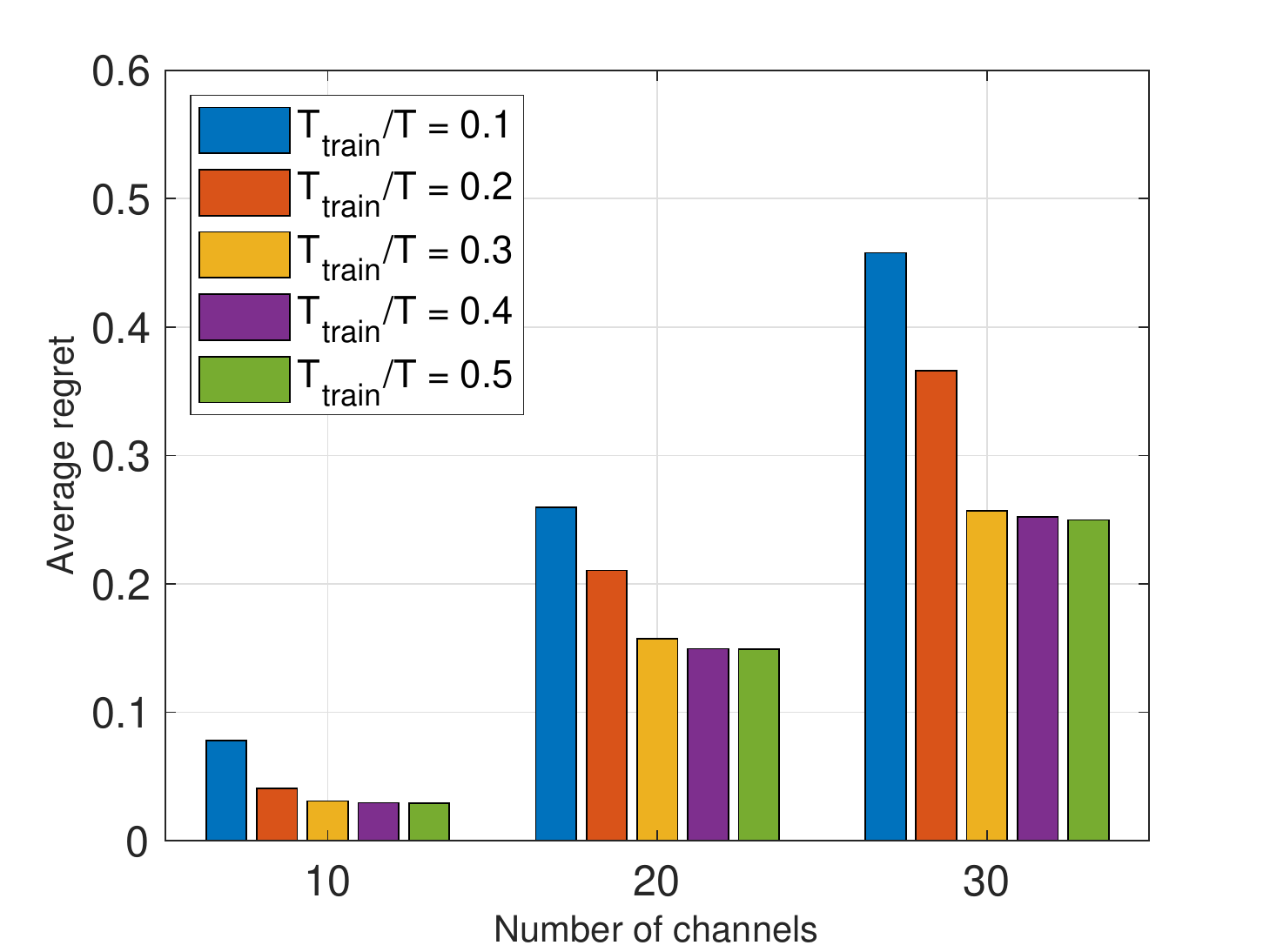}
\caption{TS}
\label{fig_regret_TS}
\end{subfigure}
\caption{Average regret vs. train time length}
\label{fig_regret}
\end{figure}

\subsubsection{Regret}
Fig. \ref{fig_regret} demonstrates the regret according to the number of channels, the length of training period, and the method of RL---viz. $\epsilon$-greedy or TS. Comparing Figs. \ref{fig_regret_greedy} and \ref{fig_regret_TS} suggests that TS results in a smaller regret as compared to $\epsilon$-greedy. The reason is that TS wastes a smaller number of trials for exploring arms with lower chances of winning than $\epsilon$-greedy does. Now, within each of Figs. \ref{fig_regret_greedy} and \ref{fig_regret_TS}, it is apparent that the regret is elevated with (i) a shorter training period and (ii) a larger number of channels to explore.

\section{Related Work}\label{sec_related}
\subsubsection{ITS Literature}
The literature of ITS has discussed the significance of security and privacy in vehicular networks \cite{tits_survey}. A critical challenge in achieving security in a V2X network is the complexity and dynamicity attributed to mobility. A recent work in the literature proposed an optimal decision algorithm that is able to maximize the chance of making a correct decision on the message content, assuming the prior knowledge of the percentage of malicious vehicles in the network \cite{tits_dissemination}.

Meanwhile, the literature has also discussed blockchain as a main technological component to promote trust among vehicles. As an example, a privacy-improving blockchain architecture has been proposed \cite{tits_creditcoin}. The mechanism features a blockchain mechanism enabling signatures to be exchanged without revealing the sender's identity, as a means to improve privacy.

However, none of the prior work has adequately addressed the key issue that this paper targets to discuss: the scalability for blockchain applied to a V2X network.

\subsubsection{Scalability Trilemma}
Permissioned blockchains such as Hyperledger Fabric \cite{fabric_ibm18} and Zyzzyva \cite{zyzzyva} employ speculative consensus methods, which increase scalability under the assumption that the security can be kept as long as $3f+1$ nodes participate in a consensus (where $f$ is the number of faults). A key limitation from the literature is that no further details were discussed about optimization of selection of voting peers.

\subsubsection{Blockchain-Empowered Networks}
We found a body of prior work discussing blockchain applied on vehicular networks. An example is a RL-based industrial internet of things (IoT) \cite{icc19_tii19}. Another RL-based performance optimization framework for blockchain-enabled internet of vehicles (IoV) was found, where transactional throughput is maximized while guaranteeing the decentralization, latency and security of the underlying blockchain system \cite{acm18}. However, these existing methods limit its own applicability by assuming that a vehicle is able to select a certain consensus method. In practice, it is hard to switch the blockchain parameters in the middle of operating a consensus procedure.

Another proposal focused on the endorsement procedure of Fabric \cite{india19}. Revealing the identity of an endorser to the peer nodes may not be suitable for transactions in which the endorsing peers have different preferences. However, it shows a limitation: in V2X, not every transaction will require anonymous endorsement. Thus, this proposal lacks the generality.

\subsubsection{Improvement of Hyperledger Fabric}
In the current version of Fabric, a client could only \textit{guess} in a selection of endorsing peers for a transaction \cite{fabric_ibm18}. Implementation was not dynamically reactive to network changes (such as the addition of peers who have installed the relevant chaincode, or peers that are temporarily offline). Static configurations also do not allow applications to react to changes of the endorsement policy itself (as might happen when a new organization joins a channel). Furthermore, the client application had no way of knowing which peers have updated ledgers and which do not, so it might submit proposals to peers whose ledger data is not in sync with the rest of the network, resulting in transaction being invalidated upon commit. That was a waste of both time and resources. V2X is dynamic to rely on this probabilistic method. Fast processing is needed while keeping ``liveness.''

As a remedy, the Fabric recently added the \textit{service discovery} \cite{ibm19}. But it comes at the cost of higher complexity due to the need for additional information to each client. A scalability issue is anticipated with a very large number of clients. Higher security threat to malicious clients masquerading legitimate ones. Also, another proposal suggested an \textit{anonymity} of endorsing peers in order to prevent a bias \cite{india19}. However, not every application is biased; thus, it may incur unnecessary inefficiency if an application does not need anonymity.

\section{Conclusions}
This paper proposed a RL-based channel selection framework for the Hyperledger Fabric applied to V2X networks. We formulated the machine learning as a contextual MAB problem with the length of a vehicle's dwelling time in a Fabric network as the context. Specifically, we found that a tradeoff exists on the number of peers in a channel: a procedure of endorsement and consensus becomes (i) less scalable with too many peers and (ii) susceptible to faults with too few peers. Also, since the vehicle has no prior information of the peers' probability of fault upon joining a network, there is no way to anticipate the performance of each channel until it has learned about it. As an actual means to perform the learning, the proposed framework enabled a vehicle to adopt $\epsilon$-greedy and or TS. The results of our experiments showed that the proposed RL mechanism led to stable selection of channels fulfilling the success condition. More precisely, the proposed algorithm showed the latency and throughput close to the optimal.

This work is expected to have significant impact on future applications across the technologies gaining high research interest, namely Hyperledger Fabric and V2X. Despite its unique modularized, execute-order structure, the Hyperledger system still has many aspects unproven when applied to V2X. One possible extension of this work is to incorporate the proposed RL mechanism to incorporate other dynamic factors such as network condition and evaluate the resulting performance impacts.

\section*{Acknowledgement}
The work of Ahmed S. Ibrahim is supported in part by the National Science Foundation under Award No. CNS-1816112.


\end{document}